\documentclass[runningheads]{llncs}
\usepackage[T1]{fontenc}
\usepackage{graphicx}
\usepackage{comment}
\usepackage{algpseudocode}
\usepackage{amsmath}
\usepackage[margin=3cm]{geometry}
\usepackage[ruled,vlined,linesnumbered,noresetcount]{algorithm2e}

\begin{document}
\def\thefootnote{$*$}\footnotetext{These authors contributed equally to this work}

\title{Data-Driven Approach to assess and identify gaps in healthcare set up in South Asia}
%
\author{R. Elahi\inst{1}\textsuperscript{$*$} \and Z. Tahseen\inst{2}\textsuperscript{$*$} \and T. Fatima\inst{1}\textsuperscript{$*$} \and S. W. Zahra\inst{1} \and H. M. Abubakar\inst{1} \and T. Zafar\inst{3} \and A. Younas\inst{4} \and M. T. Quddoos\inst{5}\textsuperscript{$*$}  \and U. Nazir\inst{1}\textsuperscript{$*$}}
\authorrunning{U. Nazir et al.}

\institute{Center for AI Research (CAIR), School of Computer and Information Technoglogy, Beaconhouse National University, Pakistan,\\
\email{\{f2021-711, f2021-557, f2021-136, f2021-641, usman.nazir\}@bnu.edu.pk}
\and 4XPillars\\
\email{z.tahseen@4xpillars.com}
\and Allama Iqbal Medical College\\
\email{tehreemzafar687@gmail.com}
\and 46 Labs\\
\email{aqs.younas@46lab.com}
\and Center for Urban Informatics, Technology and Policy (CITY) at LUMS \\
\email{muhammad.quddoos@lums.edu.pk}
}

\maketitle

\begin{abstract}
Primary healthcare is a crucial strategy for achieving universal health coverage. South Asian countries are working to improve their primary healthcare system through their country specific policies
designed in line with WHO health system framework using the six thematic pillars: Health Financing, Health Service delivery, Human Resource for Health, Health Information Systems, Governance, Essential Medicines and Technology,  and an addition area of Cross-Sectoral Linkages \cite{indicators2010monitoring}. Measuring the current accessibility of healthcare facilities and workforce availability is essential for improving healthcare standards and achieving universal health coverage in developing countries. Data-driven surveillance approaches are required that can provide rapid, reliable, and geographically scalable solutions to understand a) which communities and areas are most at risk of inequitable access and when, b) what barriers to health access exist, and c) how they can be overcome in ways tailored to the specific challenges faced by individual communities. We propose to harness current breakthroughs in Earth-observation (EO) technology, which provide the ability to generate accurate, up-to-date, publicly accessible, and reliable data, which is necessary for equitable access planning and resource allocation to ensure that vaccines, and other interventions reach everyone, particularly those in greatest need, during normal and crisis times.
This requires collaboration among countries to identify evidence based solutions to shape health policy and interventions, and drive innovations and research in the region.

\end{abstract}

\section{Introduction}


\begin{figure}[h]
    \centering
    \includegraphics[width=1\textwidth, trim={0cm 2.5cm 0 0},clip]{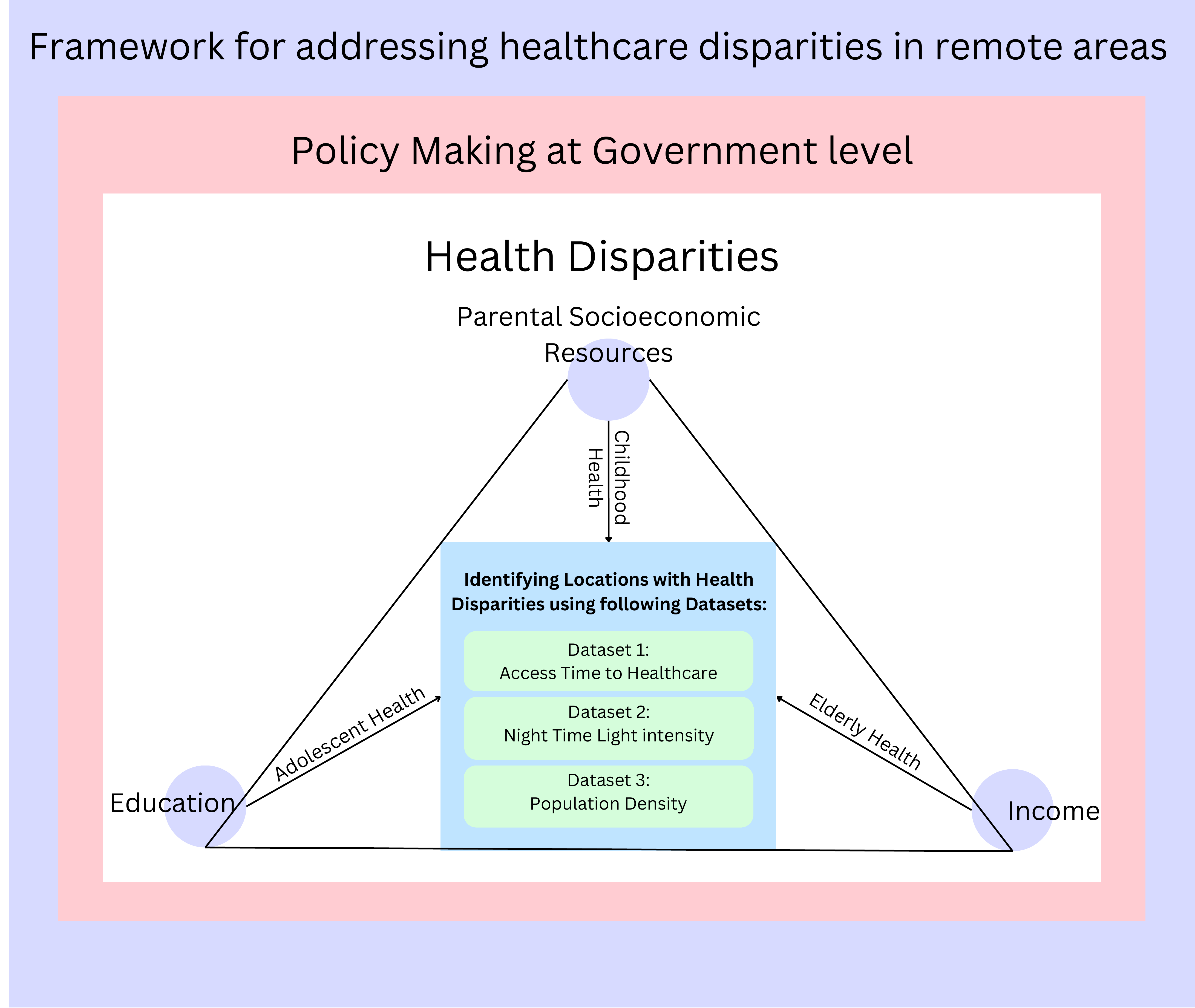}
    \caption{Health Disparities due to a) lack of access to infrastructure, e.g., health facilities, through interventions, e.g., safe medicines and vaccines;  b) access to communication infrastructure in remote locations, e.g., poor road networks and communication links;  and c) affordability and intersectional determinants of access, such as gender, age, and lifestyle factors.}
    \label{fig:healthdisparity}
\end{figure}

Even while the Sustainable Development Goal 3 (SDG 3) is being achieved, 381 million people, or 4.9\% of the world's population, still live in severe poverty and do not have access to even the most basic healthcare~\cite{united2023sustainable,sachs2022sustainable}. Mortality rates are highest in sub-Saharan Africa and Southern Asia, essentially due to insufficient access to primary healthcare services ~\cite{united2023sustainable}. It is incumbent to address discrepancies in order to close this gap and guarantee equitable access to healthcare. To achieve the vision of “Health for All”, we must address the systemic barriers that perpetuate health inequities \cite{world2015mdgs,carlsen202217}. 
Inequity has many dimensions \cite{statistics2017monitring} as shown in figure~\ref{fig:healthdisparity}; barriers include and are not limited to: a) lack of access to infrastructure, e.g., health facilities
\cite{tichenor2021interrogating};  b) access to communication infrastructure in remote locations, e.g., poor road networks and communication links \cite{international2020measuring};  and c) affordability and intersectional determinants of access, such as gender, age, and lifestyle factors \cite{world2021addressing}. 

Insufficient and ineffective healthcare methods exacerbate the lack of appropriate access to healthcare \cite{hazell2006under}, and many low- and middle-income nations lack real-world health data infrastructure due to overworked healthcare systems. In heavily populated areas, particularly urban slums and remote rural areas. Manual monitoring and data gathering in remote locations can be unacceptably time-consuming, costly, prone to human error and malpractice and unsustainable given limited resources \cite{kiberu2014strengthening}. 

Data-driven surveillance approaches are required that can provide rapid, reliable, and geographically scalable solutions to understand: a) which communities and areas are most at risk of inequitable access and when \cite{ibrahim2021health}; b) what barriers to health access exist; and c) how these barriers can be overcome in ways tailored to the specific challenges faced by individual communities \cite{marmot2012european}. It can help to reduce health data poverty and disparities. Better knowledge, achieved through better data and robust decision-support systems, can help address everyone's health requirements in an equitable manner \cite{committee2003future}.

The goal of the Centers for Disease Control and Prevention's (CDC) Data Modernization Initiative (DMI) is to increase the effectiveness and efficiency of health data collecting, sharing, and analysis. On the other hand, there isn't much research on integrating social determinants of health (SDOH) into state and federal DMI initiatives. The terms SDOH refer to an individual's living, learning, and working environments, as well as the ways in which these environments impact their health. Examples of SDOH include safe housing, access to wholesome food, employment possibilities, and transportation. 

We propose to harness current breakthroughs in Earth-observation (EO) technology, which provides the ability to generate accurate, up-to-date, publicly accessible, and reliable data, which is required for equitable access planning and resource allocation to ensure that safe medicines, vaccines, and other interventions reach everyone, particularly those in greatest need, during normal times  \cite{topol2019high,citaristi2022united}. This data can also be used in emergency scenarios such as pandemics and natural catastrophes, which disproportionately affect underserved groups~\cite{nazir2024predicting}. Therefore, this data creation can help identify requirements and track progress towards increasing equal access to healthcare worldwide. This is particularly important in resource-constrained areas, both in normal and emergency situations  \cite{world2018world}. Figure.~\ref{fig:healthdisparity} illustrates how government-level policies can address health inequalities, a key focus of SDG 3 ~\cite{united2023sustainable,sachs2022sustainable}. It emphasizes the role of parents' socioeconomic resources in influencing children's health, which in turn influences the health of the young and old, as well as education and income levels. Our approach can identify health gaps in peripheral areas and improve surveillance, especially in resource-limited areas. By harnessing the power of EO  and artificial intelligence, we can create a fairer and more sustainable healthcare system that leaves no one behind.  

Our major contributions to mapping health equities in South Asia, as compared to AccessMod 5 (see Appendix \ref{accessmod}), include the following:
\begin{itemize}
    \item Remote areas and potential improvement neighborhoods with missing nighttime light intensity: We use nighttime light intensity data to find unmet needs in the most remote areas. With this method we can also identify geographical areas with poorly served health services in terms of infrastructure development 
    \item Localizing health disparities in underserved remote areas: We focus on pinpointing specific health disparities within the identified under-served areas. This localization helps to understand the specific health needs and challenges faced by these communities, allowing for targeted interventions.
\end{itemize}

\begin{figure}
    \centering
    \includegraphics[scale=0.4, trim={0cm 2.5cm 0 0},clip]{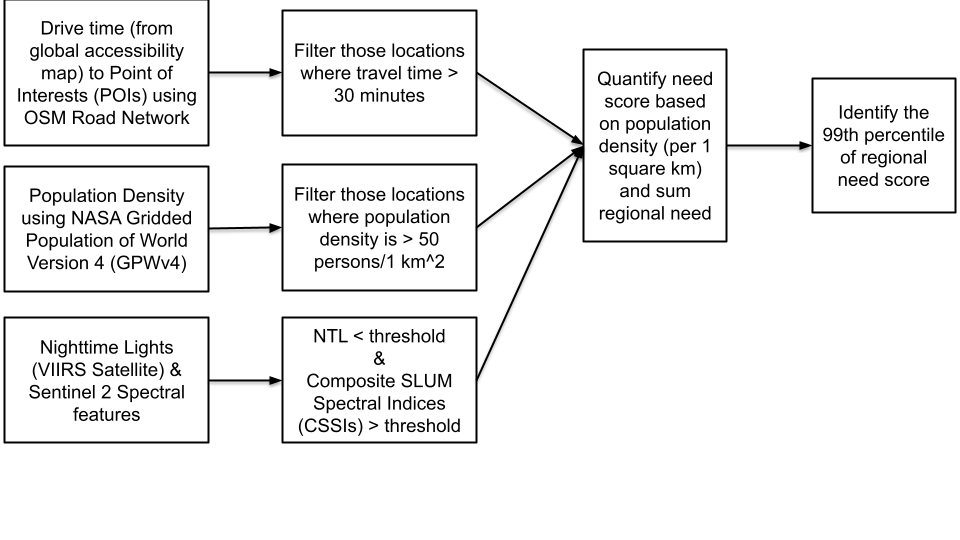}
    \caption{Proposed approach for mapping health inequities.}
    \label{fig:method}
\end{figure}

\section{Mapping Primary Health Care Infrastructure Gaps}
Strategists, academics, and public health practitioners strive to improve the overall health of the population by addressing health disparities based on geographic origin, race/ethnicity, socioeconomic status (SES), and   other social factors. Health inequalities refer to systematic variations in health status among various socioeconomic classes in a society, which are socially produced, theoretically avoidable, and often considered undesirable in a civilized society \cite{whitehead2007typology}. The concept of health inequality makes no moral judgments about whether observed disparities are equitable or right. A health inequity, also known as a health disparity, is a subtype of health inequality that refers to an unjust discrepancy in health status. Allowing health inequalities to persist is unjust, according to one common definition \cite{whitehead1992concepts}. In this sense, health inequities are systematic differences in health that could be avoided by reasonable means \cite{marmot2012european}. Social group differences in health, such as those based on race or religion, are considered health inequities because they reflect an unfair distribution of health risks and resources. The key distinction between inequality and inequity is that the former is simply a dimensional description employed whenever quantities are unequal, while the latter requires passing a moral judgment that the inequality is wrong. For example, the term "health inequality" can reflect racial/ethnic differences in US infant mortality rates, which are roughly three times greater for non-Hispanic blacks than whites \cite{people2013conclusion}.

Age-related health disparities, such as those between the ages of 20 and 60, are seen as health inequalities and are perceived as unavoidable rather than unjust. By contrast, disparities in health are investigated among populations worldwide according to factors such as income, caste, gender, education, race/ethnicity, and employment. To investigate these disparities, researchers compare socioeconomic variables and examine variations in health outcomes at the group level. For example, comparing the mean body mass index (BMI) of the wealthy to that of the impoverished can help explain socioeconomic differences.
 In order to reduce disparities brought about by the unequal distribution of social determinants of health, policy development and resource allocation must acknowledge and address social group health inequalities \cite{arcaya2015inequalities}. 
The World Health Organization proposes that health indicators be reported by groups, or equity stratifiers, in order to monitor health disparities. By concentrating on social groupings, we may comprehend present health inequalities within a historical and cultural framework and subsequently offer an explanation of the mechanisms underlying health inequality  \cite{world2013handbook}. For instance, knowledge of American racism and slavery past can help explain current racial/ethnic health disparities. Similarly, studying the political and theological history of India's caste system allows us to better appreciate how it influences people's social position, employment, education levels, and health outcomes. Viewing health inequalities through the lens of social groups can help direct actions, enable surveillance of key equity concerns, and increase our knowledge of health by making connections that were not initially clear. Health inequalities along racial, ethnic, and socioeconomic lines exist in both low- and high-income nations, and they may be expanding, emphasizing the necessity of investigating group-level health discrepancies \cite{braveman2002social}. Understanding socially patterned health inequalities requires defining meaningful social groups. Each community has its own distinct method of classifying and separating people into social divisions.

The second related topic is whether absolute or relative position matters for health. This is especially relevant when it comes to poverty, which can be defined in two ways: absolute (comparing a given income to a static benchmark) and relative (comparing a given income to the general distribution of incomes in a community). Absolute poverty definitions are based on a set monetary barrier known as a
poverty line, though this barrier is normally determined by the year, nation, and household size. Those with incomes less than the criteria are deemed destitute. In contrast, relative poverty is defined by comparing a particular income to the income distribution in a population. For example, one could argue that a person is comparatively impoverished if their income is less than thirty percent of the national per capital income  \cite{arcaya2015inequalities}. This suggests that the concept of poverty is not static and may change over time. Wealth, income, and education are a few factors that contribute to an individual's socioeconomic position (SEP). Education can be defined as the highest level attained or the total number of years spent in school. The main factor affecting the material resources component of SEP is income, which is often calculated as household gross income divided by the number of participants. Wealth is defined as income plus all acquired material resources.

Occupation-based indicators are also often utilized. For example, the Registrar General's Social Classes classify jobs according to prestige into six hierarchical groupings, which are frequently classified as manual vs. non-manual. The Erikson and Goldthorpe Class Schema classifies jobs according to employment relations features, while the UK National Statistics Socio-Economic Classification follows similar concepts. Other categories, which concentrate on social interaction patterns within occupational groupings, include the Cambridge Social Interaction and Stratification Scale and Wright's Social Class Scheme. The following are used to illustrate societal patterns: housing issues, overpopulation, unemployment, and composite or proxy indicators. Geographic environment, not simply social group, influences health. Health inequalities can be impacted by people's physical and social settings. Geographic health inequities may be monitored by comparing health outcomes across various areas, taking into account both place (particular geographic locations) and space (broader geographical settings) \cite{macintyre2000ecological,jones1993medical}. Researchers must select how to present observed differences while tracking health disparities across time. Inequalities between groups can be expressed as absolute or relative differences. Absolute differences represent the actual disparity in health outcomes between groups, whereas relative differences assess the proportional gap in relation to the baseline health state \cite{oliver2002addressing,king2012use}.

Our proposed technique for mapping health disparities uses a variety of data sources and approaches to quantify and identify areas with severe health needs. As shown in the flowchart (Fig. 1), we employ travel time from global accessibility maps to places of interest (POIs), population density data from NASA’s Gridded Population of the World Version 4 (GPWv4), and nighttime light data from VIIRS satellites and Sentinel 2 spectral characteristics. This method selects places based on travel time (>30 minutes), population density (>50 per km²), and thresholds for nighttime lights and composite SLUM spectral indices. By measuring regional need ratings based on population density and aggregating them, we can discover the 99th percentile of areas with severe health needs. 

This strategy is based on data that support established patterns of population and geographic health disparities. This enables the development of targeted interventions and policies to address health inequalities. Our approach provides a comprehensive framework for monitoring and addressing health inequities both at scale and at scale, combining multiple data sources with a regional and geographic focus. \textbackslash{}OF.

The rest of the paper is organized as follows: Section 3 describes how to implement the proposed approach, including the integration of open source datasets such as VIIRS NTL data, GPWv4 and Accessibility to Healthcare 2019. It also presents the analytical findings and visualizations that resulted from these datasets. Section 5 provides an overview of the results of our analysis and highlights the urgent need to address global health disparities and the value of our data strategy to guide targeted action. Chapter 6 describes actions to address health disparities, including a multifaceted strategy that combines community-based solutions with modern data analytics and earth observation technology. Section 4 concludes the paper with a summary of the main conclusions, results and evaluation. Finally, Section 7 discusses how policies can be implemented to reduce health inequalities. It supports evidence-based interventions that support equitable health outcomes and access worldwide.


\begin{algorithm}
\scriptsize
\caption{Mapping Health Inequities in South Asia}
\label{alg:method}
\SetNlSty{textbf}{(}{)}
\SetAlgoNlRelativeSize{0}
\SetAlgoNlRelativeSize{1}
\SetNlSty{textbf}{[}{]}
\SetNlSty{}{}{}
\SetKwFunction{fitpoly}{fit\_poly}
\SetKwFunction{getroots}{get\_roots\_derivative}
\SetKwFunction{isempty}{is\_empty}
\SetKwFunction{prune}{prune}
\SetKwFunction{remove}{remove}
\SetKwComment{Comment}{//}{}
\SetKwData{x}{x}
\SetKwData{y}{y}
\SetKwData{f}{f}
\SetKwData{time}{time}
\SetKwData{updatedroots}{updated\_roots}
\SetKwData{th}{th}
\SetKwData{pop}{pop}
\SetKwData{povertyAreas}{povertyAreas}
\SetKwData{needScore}{needScore}
\SetKwData{needPerPixel}{needPerPixel}
\SetKwData{highNeedScore}{highNeedScore}
\SetKwData{regionalNeedScore}{regionalNeedScore}

\SetKwInOut{Input}{Input~~~}\SetKwInOut{Output}{Output}
\SetKwProg{Fn}{Function}{}

\BlankLine

\Input{datasets $\gets$ [Accessibility to Healthcare 2019], [The Gridded Population of World Version 4 (GPWv4), Revision 11], [Annual global VIIRS V2.2 nighttime lights dataset] }  
\Output{Mapping Health Inequities in South Asia}
\SetKwFunction{FMain}{Main}
\Fn{\FMain{$datasets$}:}{
    \time $\gets$ Filter locations with AccessibilityTime $>$ 30 min
    
    \pop $\gets$ Filter locations PopulationDensity $>=$ 50 $\frac{persons}{1{km}^2}$
    
    \th $\gets$ 20 \Comment*[r]{Set your desired threshold value here}

    \povertyAreas $\gets$ Filter locations with nighttime lights $<$ \th

    \needPerPixel $\gets$ Quantify need based on population density using \time, \pop \& \povertyAreas
    
    \needScore $\gets$ Sum up the need around 25 km radius around each 1 km pixel

    \highNeedScore $\gets$ Identify the ${99}^{th}$ percentile of regional need score

    \regionalNeedScore $\gets$ calculate mean regional need score of each region in ascending order

\Return{Mapping Health Inequities in South Asia}\;
}

\end{algorithm}

\begin{figure}[h]
    \centering
    \begin{tabular}{cc}
         \includegraphics[scale=0.19]{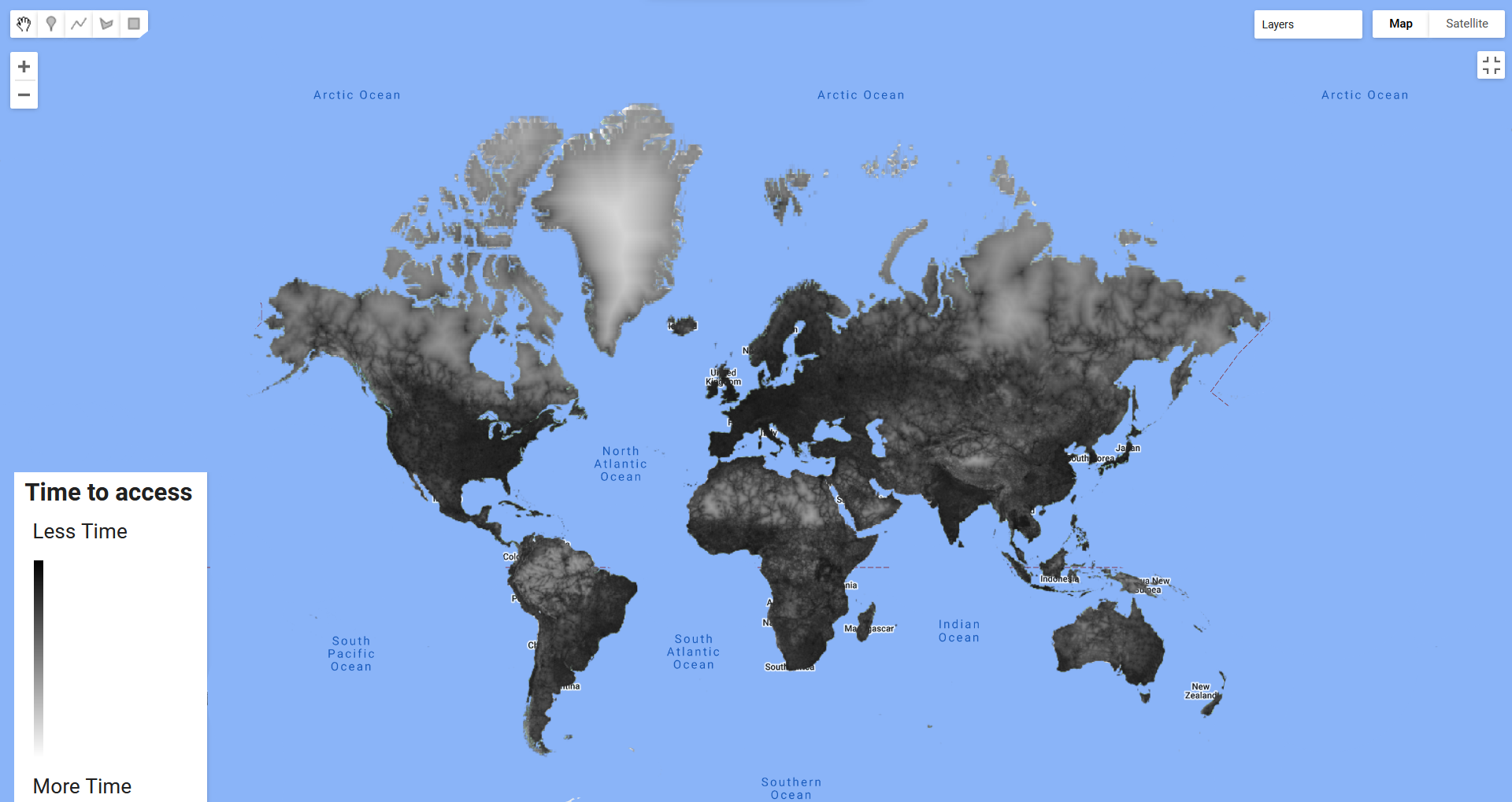} &
         \includegraphics[scale=0.19]{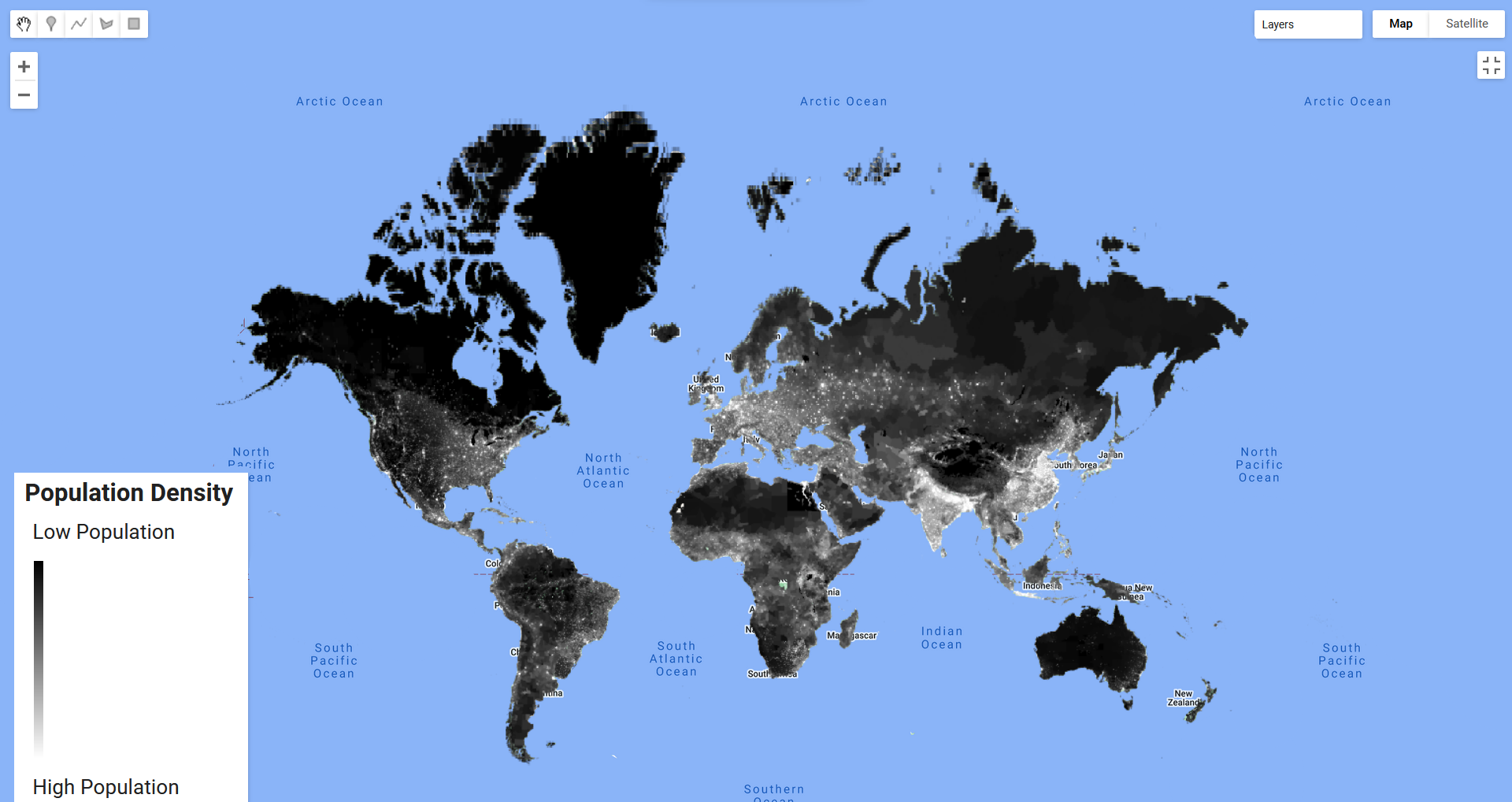}  \\
         (a) & (b) \\
         \includegraphics[scale=0.19]{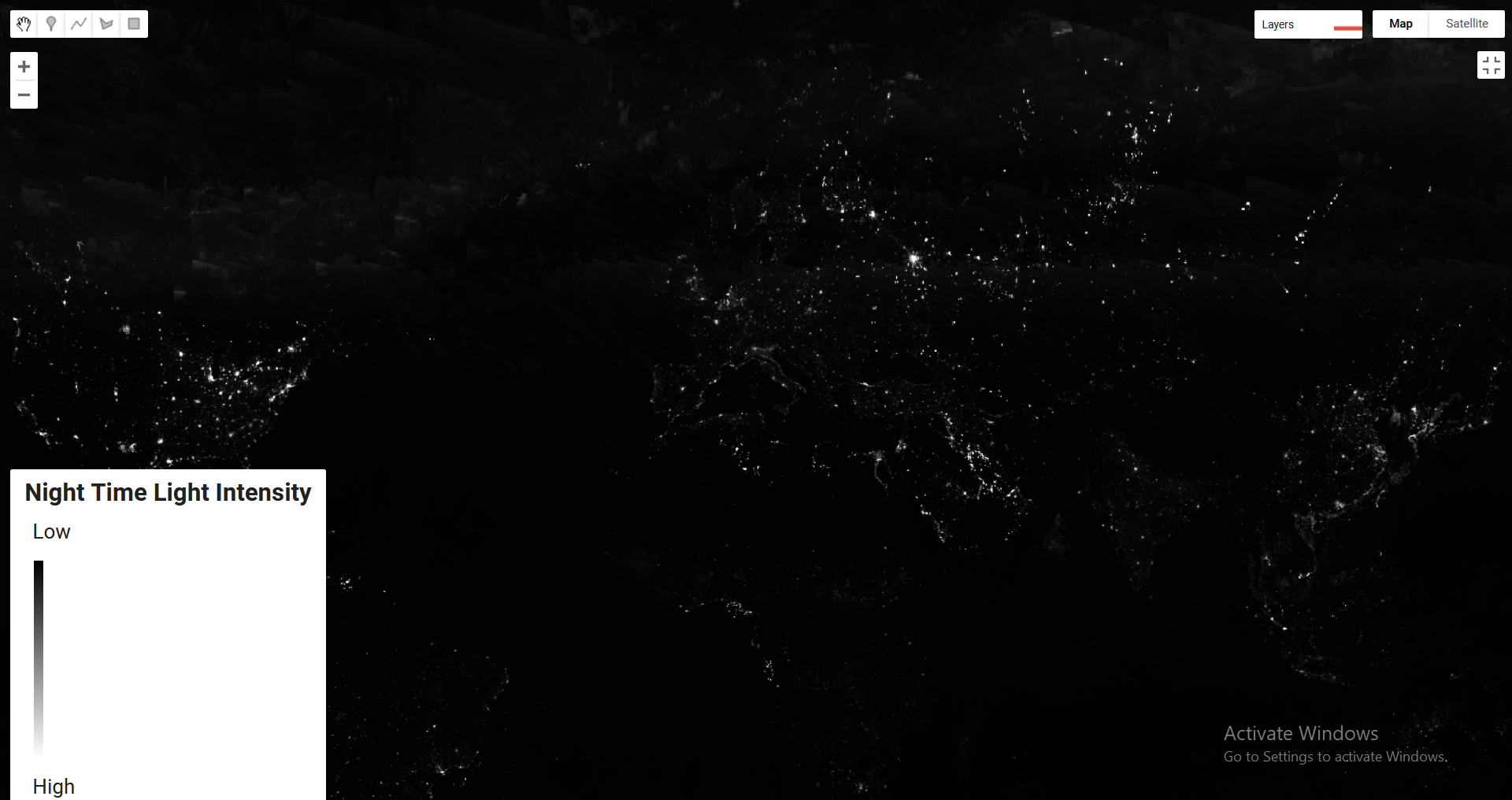} & \\
         (c) & \\
         
    \end{tabular}
    \caption{Datasets: (a): Travel time to the nearest hospital or clinic, (b): Population density using NASA Gridded Population of World version 4 (GPWv4), (c): VIIRS Nighttime Day/Night Annual Band Composites V2.2. Darker areas indicate shorter travel times to the nearest facility, lower population density, and lower nighttime light intensity compared to brighter areas.} 
    \label{fig:datasets}
\end{figure}

\section{Proposed Approach}
\subsection{Open source datasets}
The study utilizes three following primary open-source datasets (see Fig.~\ref{fig:datasets}) to identify global regions where new healthcare facilities are most needed:
\begin{enumerate}
    \item Accessibility to Healthcare 2019
    \item The Gridded Population of World Version 4 (GPWv4), Revision 11
    \item VIIRS Nighttime Day/Night Annual Band Composites V2.2 

\end{enumerate}
\subsubsection{Accessibility to Healthcare 2019}
The global accessibility map for 2019 is the result of collaboration between the University of Twente in the Netherlands, Google, Telethon Kids Institute in Perth, Australia, and MAP at the University of Oxford \cite{weiss2020global}. It shows the land-based travel time (in minutes) from any place to the nearest medical facility.  It also includes the duration of "walking-only" or using non-motorized modes of transportation exclusively. Using massive data collection efforts from Google Maps, OpenStreetMap, and university researchers, the most complete database of healthcare institution locations has been produced to date.

This project expands upon earlier research by \cite{weiss2018global}. Weiss et al. used datasets for national borders, railways, rivers, lakes, seas, topographic conditions (slope and elevation), landcover types, and roads (including the first-ever global use of Open Street Map and Google roads datasets). Each of these datasets was given a travel speed, or speeds, in terms of the amount of time needed to cross each kind of pixel. After that, the datasets were integrated to create a "friction surface," which is a map in which each pixel is given a nominal overall speed of motion according to the types that exist there. An upgraded friction surface was made for the current project to take advantage of newer developments in the OSM roads data.

\subsubsection{GPWv4.11}
The Gridded Population of World Version 4 (GPWv4), Revision 11 represents the global distribution of human population per (approximately) 1km grid cells. A proportionate distribution of the population from census and administrative units is used to distribute the population among the cells.  In terms of relative spatial distribution, these population density grids' estimations of the number of people per 1 km grid cell are in line with national censuses and population registries \cite{pagesgridded}.
\subsubsection{VIIRS Nighttime Day/Night Annual Band Composites V2.2}
 Annual global VIIRS nighttime lights dataset is a time series produced from monthly cloud-free average radiance grids for 2022 \cite{elvidge2021annual}. After removing pixels from the sun, moon, and clouds in the first filtering stage, preliminary composites with lights, flames, aurora, and backdrop were produced. To create rough annual composites, monthly increments of the rough annual composites are created and then blended.

\subsection{Finding Health Disparities}
The proposed methodology for finding health disparities as shown in figure~\ref{fig:method} and algorithm~\ref{alg:method} involve following steps:
\begin{enumerate}
    \item Importing Datasets: The analysis environment receives the datasets for accessibility, population density, and nighttime light intensity.
    \item Filtering Areas with Limited Accessibility: Locations where traveling to the nearest medical facility takes more than thirty minutes are marked. These regions are typical of those with poor access to medical care.
    \item Combining accessibility with population density and nighttime light intensity: areas with high population density (50 people per square kilometer or more), poor socioeconomic resources (night light intensity below the threshold) and poor access. to health services are identified for further study (travel time more than 30 \textbackslash{} n minutes). This policy ensures that the focus is always on densely populated areas with limited access to health services.

\end{enumerate}

\begin{figure}[h]
    \centering
    \begin{tabular}{cc}
         \includegraphics[scale=0.15]{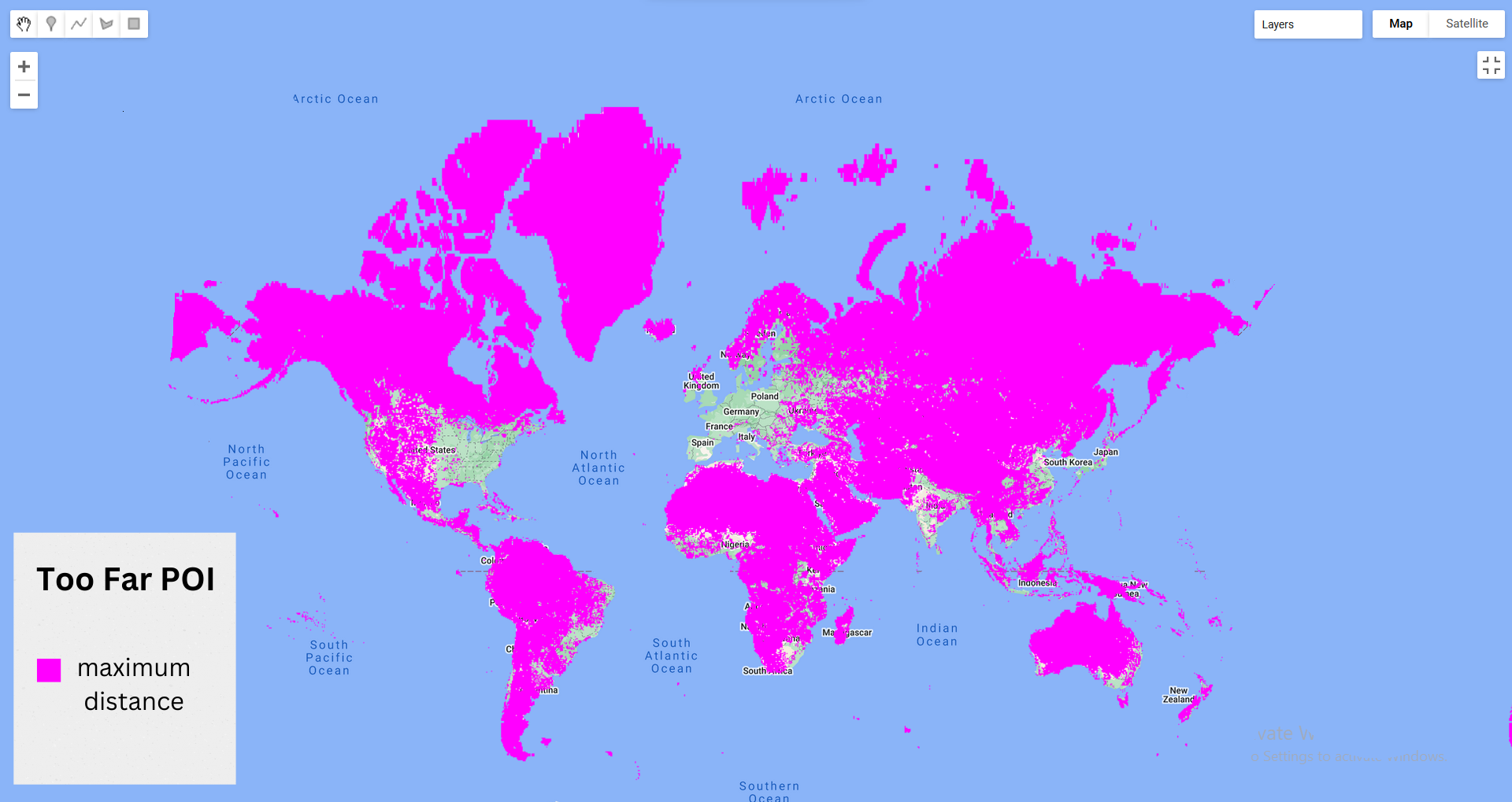} &\includegraphics[scale=0.15]{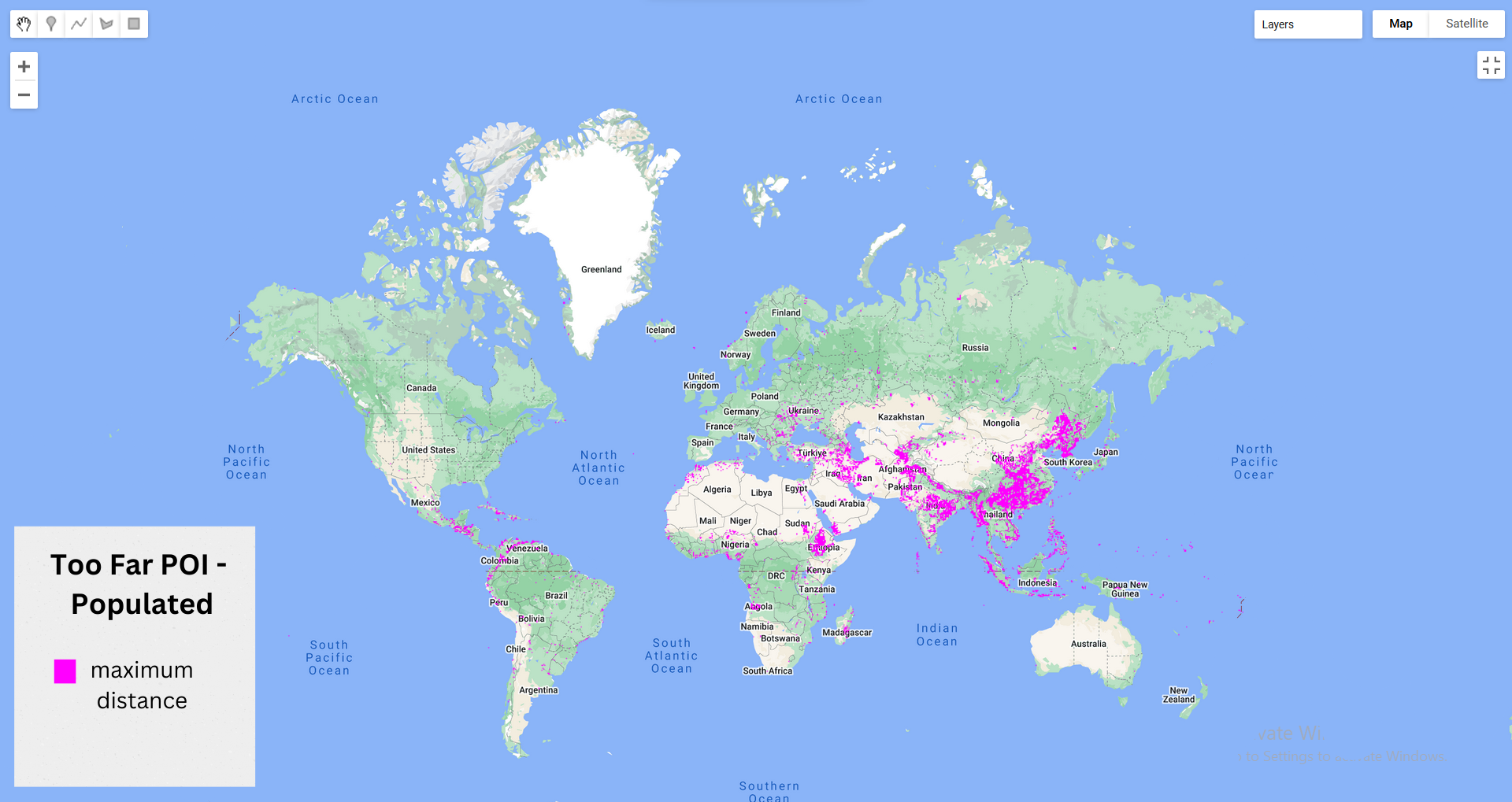} \\
         (a) & (b) \\
         \includegraphics[scale=0.15]{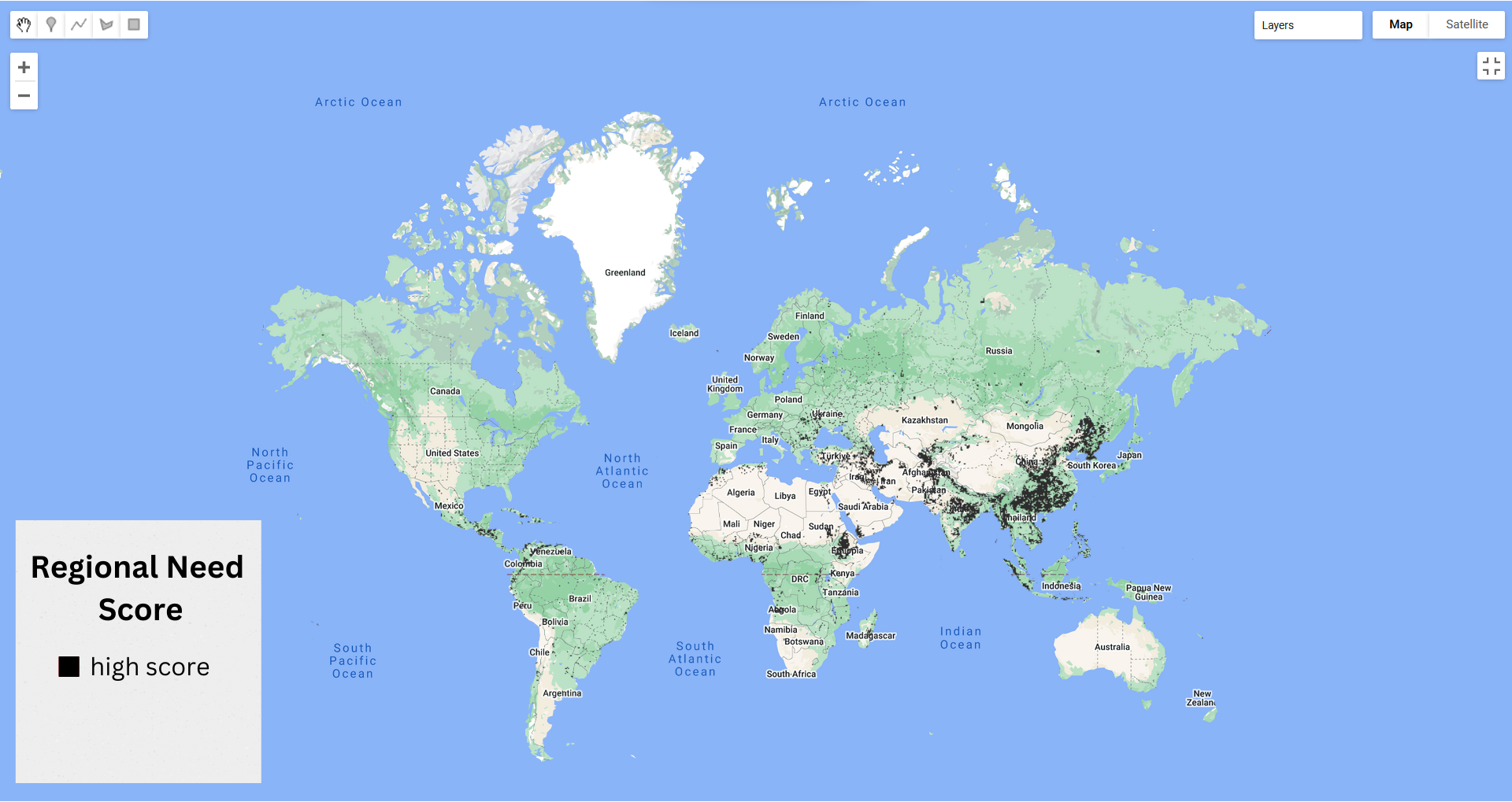} & 
         \\
         (c) &  \\
          
    \end{tabular}
    \caption{Evaluation Results: (a): Location with too far Points of interest (health facilities is greater than $30$ minutes); (b): Filter those locations where travel time is greater than $30$ minutes \& Population density is greater than $50$ persons per $1~{km}^2$; (c): Regional Need score based on population density per $1~{km}^2$ - ${99}^{th}$ percentile.}
    \label{fig:world}
\end{figure}

\begin{figure}[h]
    \begin{tabular}{ccc}
     \includegraphics[width=0.32\textwidth, trim={0.8cm 0 2cm 0},clip]{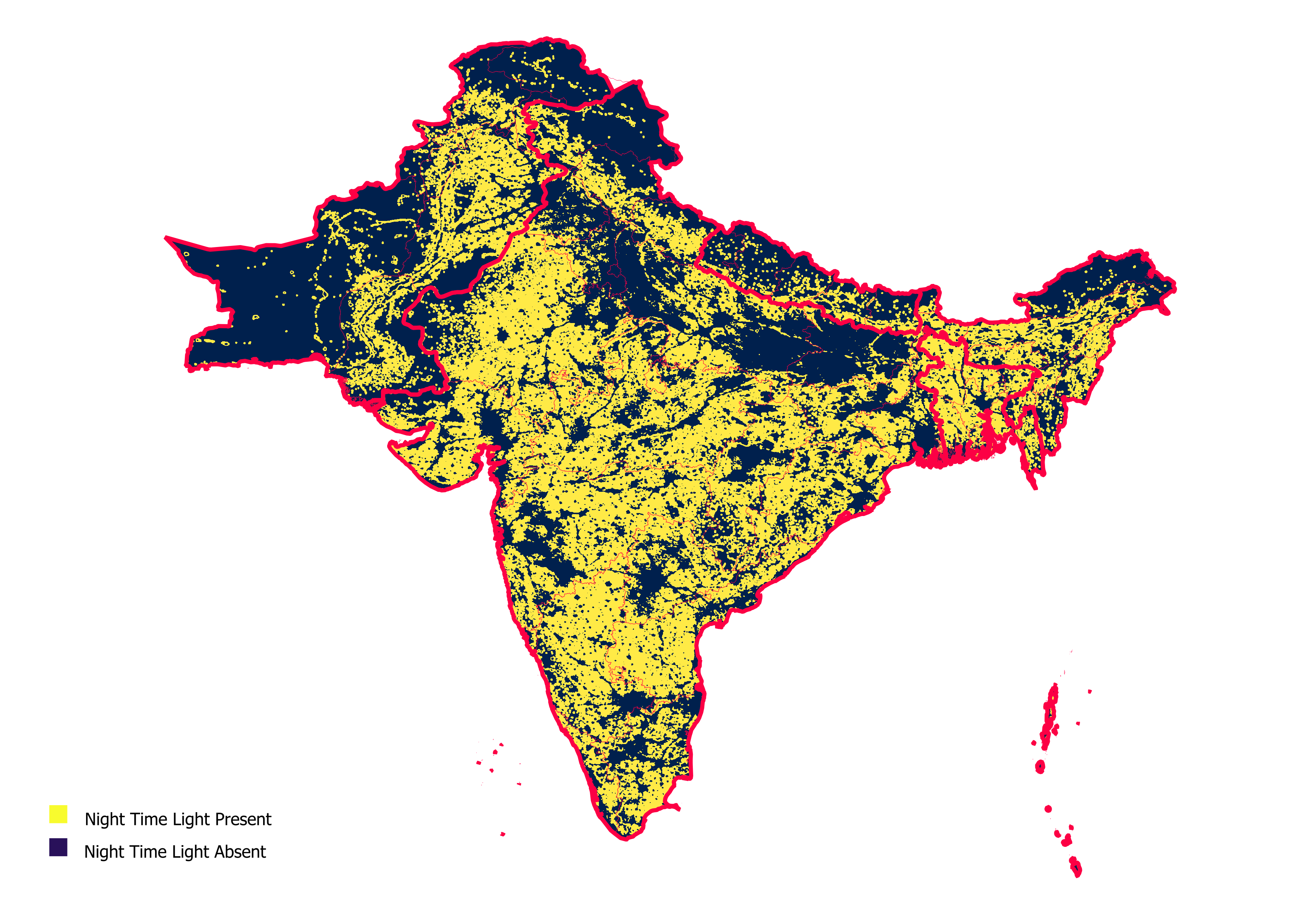}    &  \includegraphics[width=0.32\textwidth, trim={0.8cm 0 2cm 0},clip]{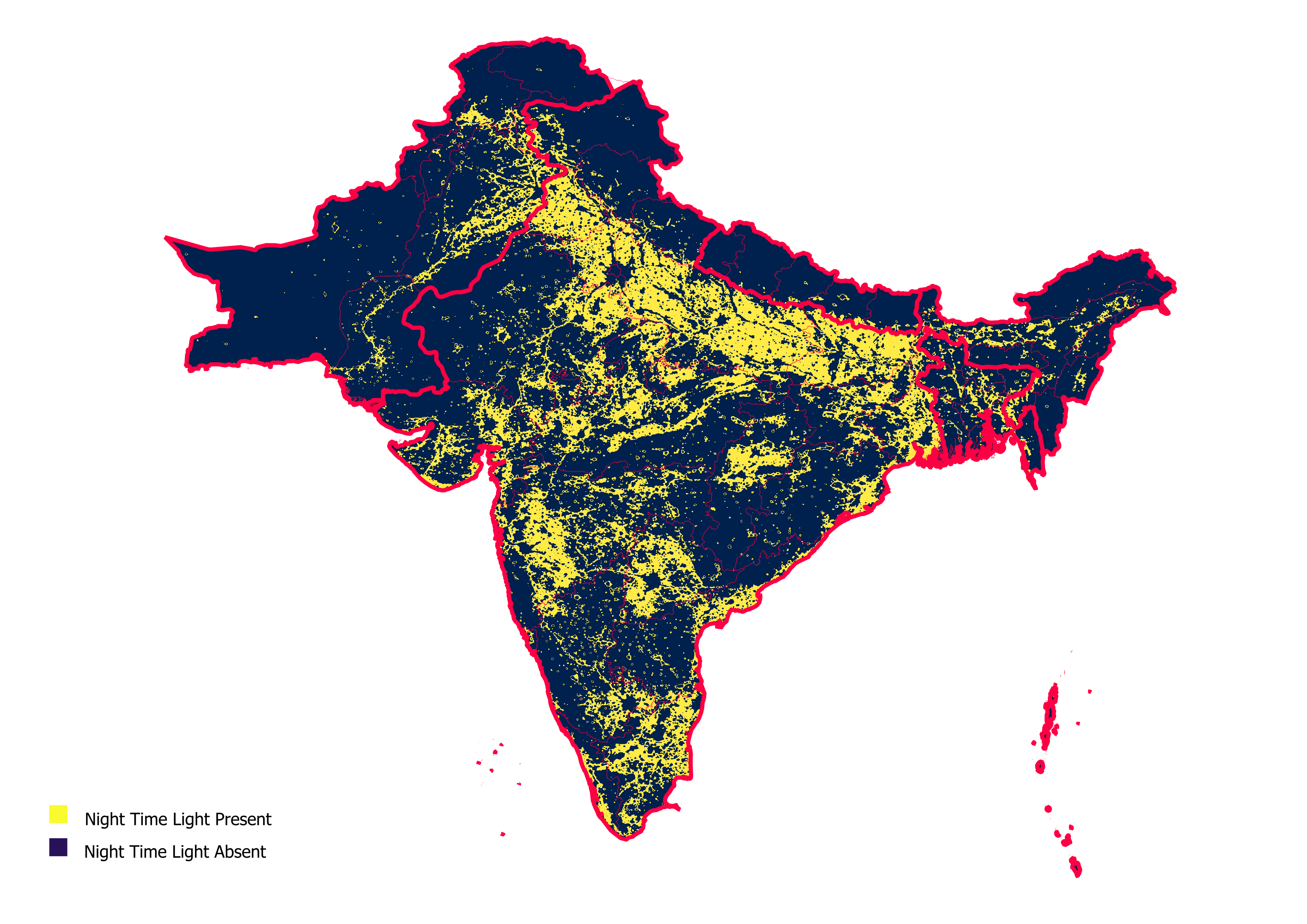}& \includegraphics[width=0.32\textwidth, trim={0.8cm 0 2cm 0},clip]{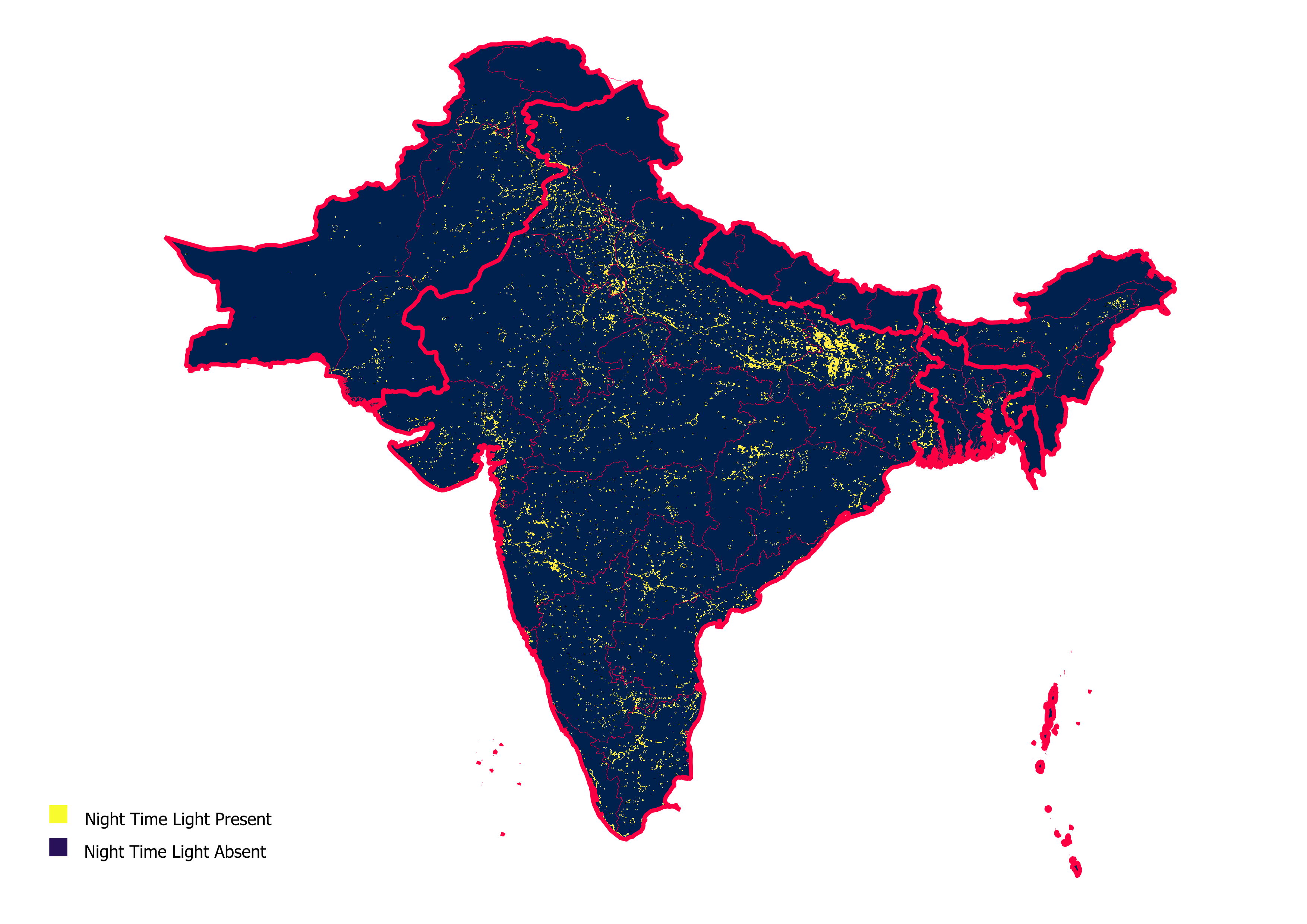} \\
     (a) & (b) & (c)\\
    \end{tabular}
    \caption{Categorization of areas based on Nighttime Light (NTL) intensity: (a) Areas with $0 \leq NTL \leq 10$; (b) Areas with $10 \leq NTL \leq 20$; (c) Areas with $20 \leq NTL \leq 30$.}
    \label{fig:ntl}
\end{figure}

\begin{figure}[h]
    \centering
    \begin{tabular}{ccc}
    \includegraphics[width=0.3\textwidth, trim={0.8cm 0 2cm 0},clip]{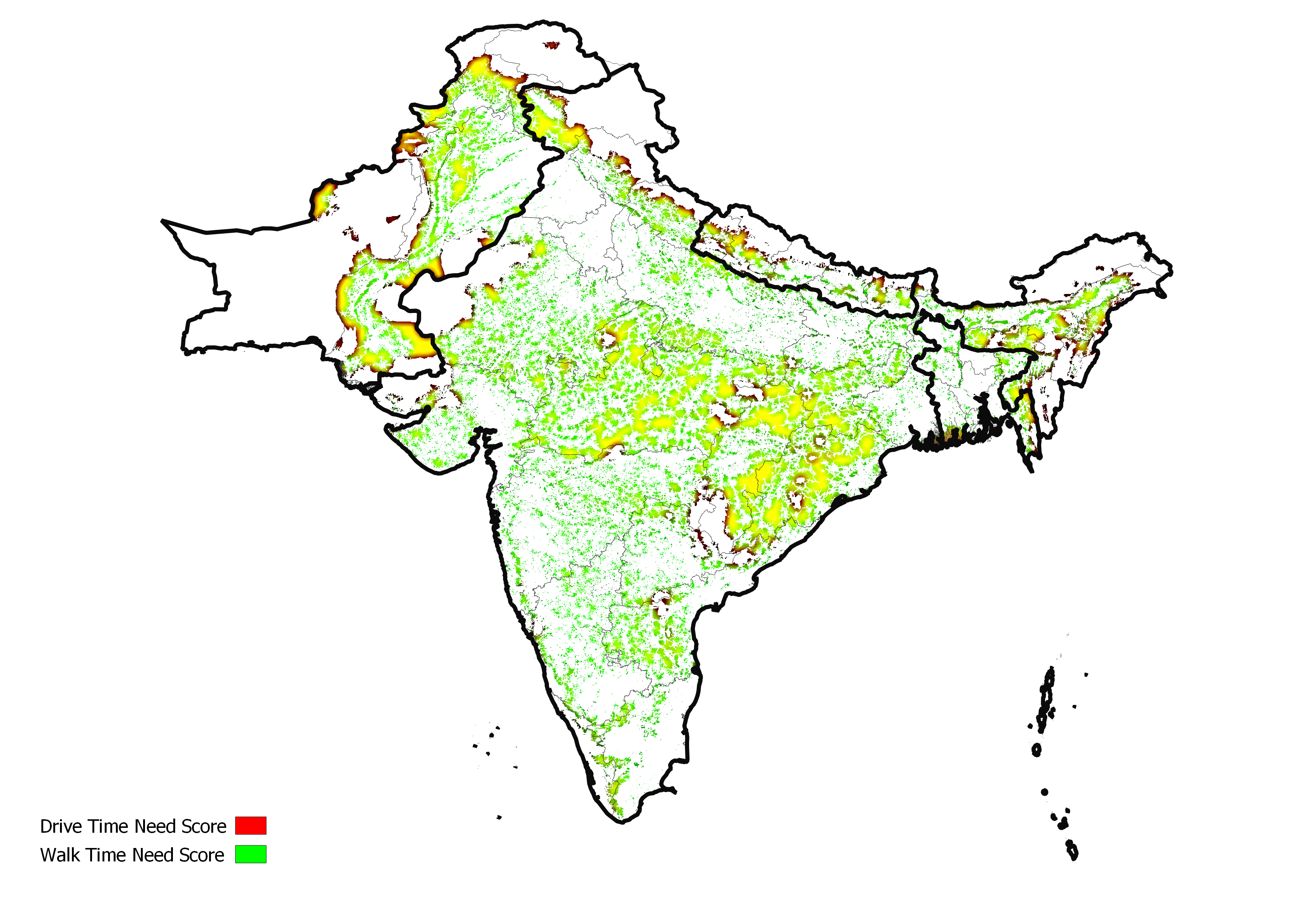} & \includegraphics[width=0.3\textwidth, trim={0.8cm 0 2cm 0},clip]{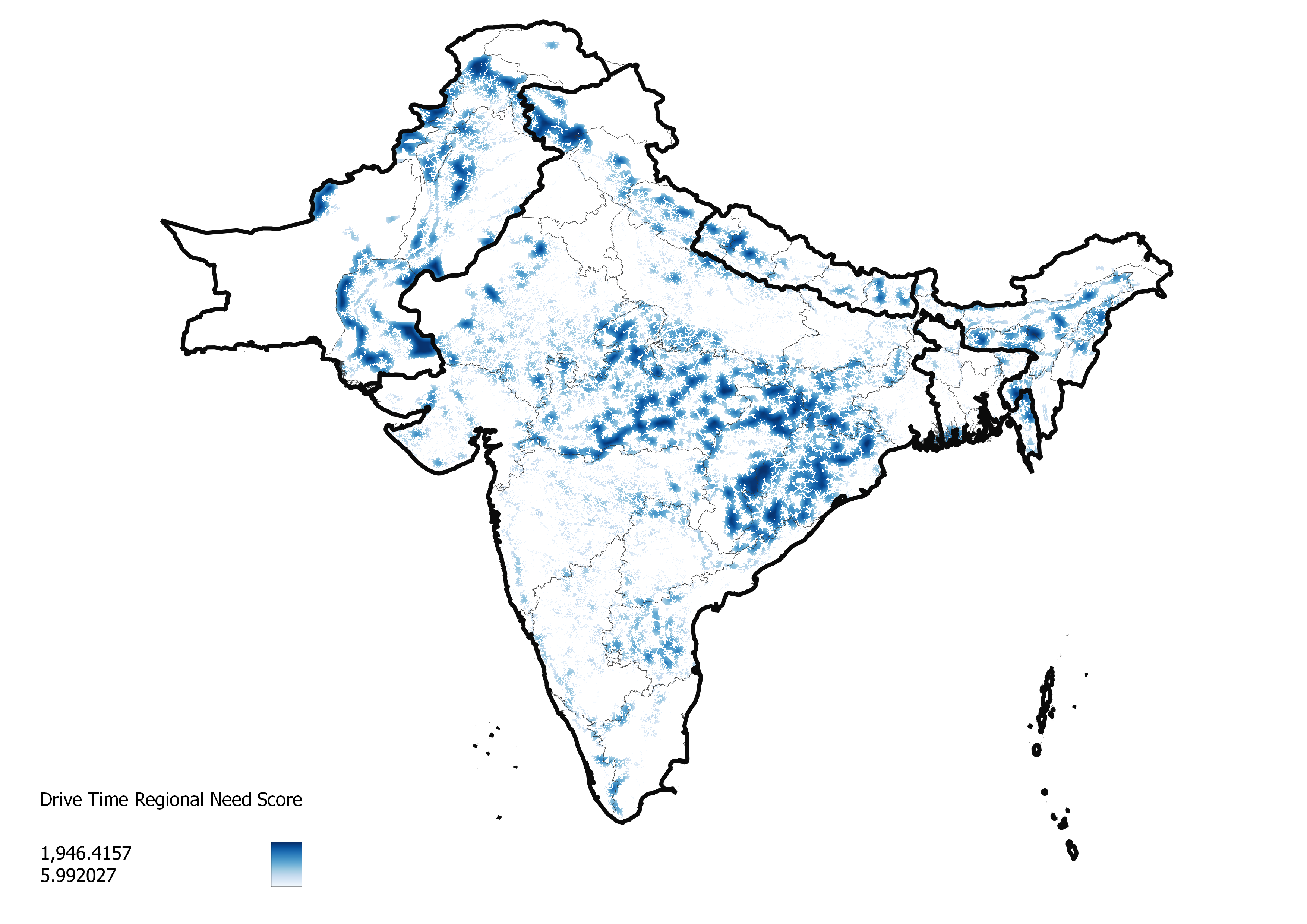} & \includegraphics[width=0.3\textwidth, trim={0.8cm 0 2cm 0},clip]{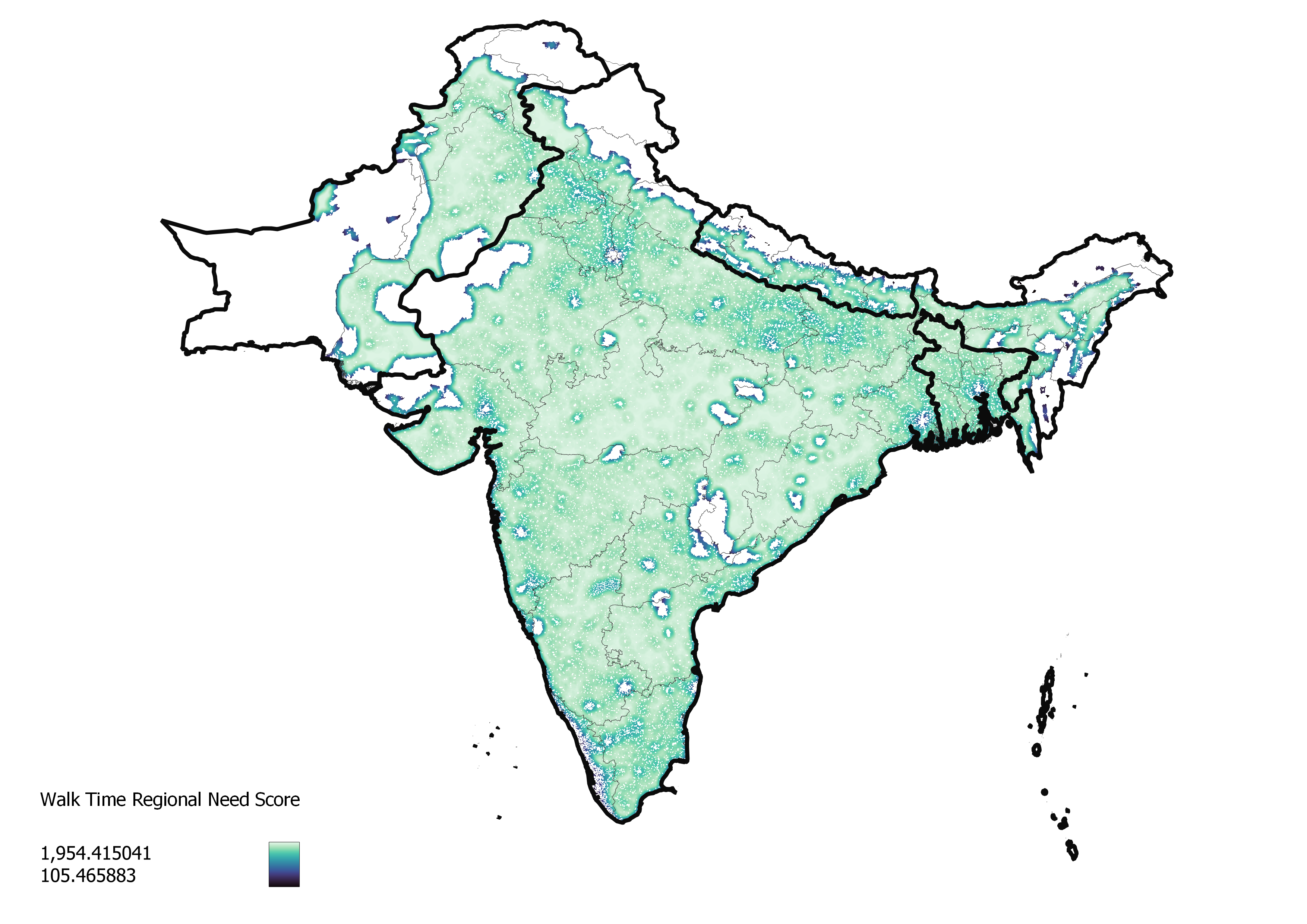} \\
    (a) & (b) & (c) \\
    \end{tabular}
    \caption{(a) Regional need score to nearest hospital/clinic based on drive time and walk time; (b) Drive time regional need score; (c) Walk time regional need score.}
    \label{fig:SouthAsiaunderserved}
\end{figure}

\begin{figure}
    \centering
    \begin{tabular}{ccc}
        \includegraphics[width=0.3\textwidth, trim={0.8cm 0 2cm 0},clip]{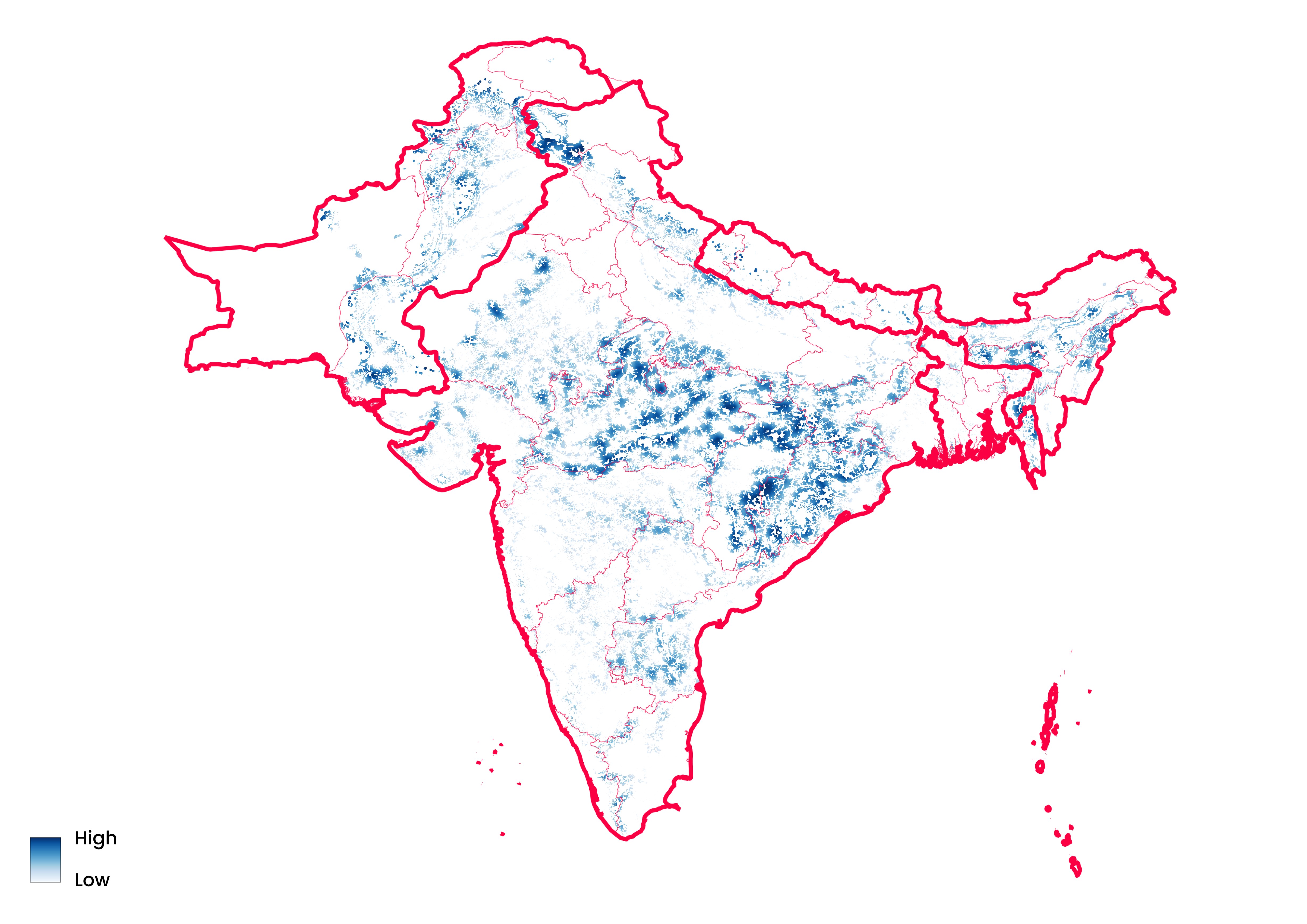} & \includegraphics[width=0.3\textwidth, trim={0.8cm 0 2cm 0},clip]{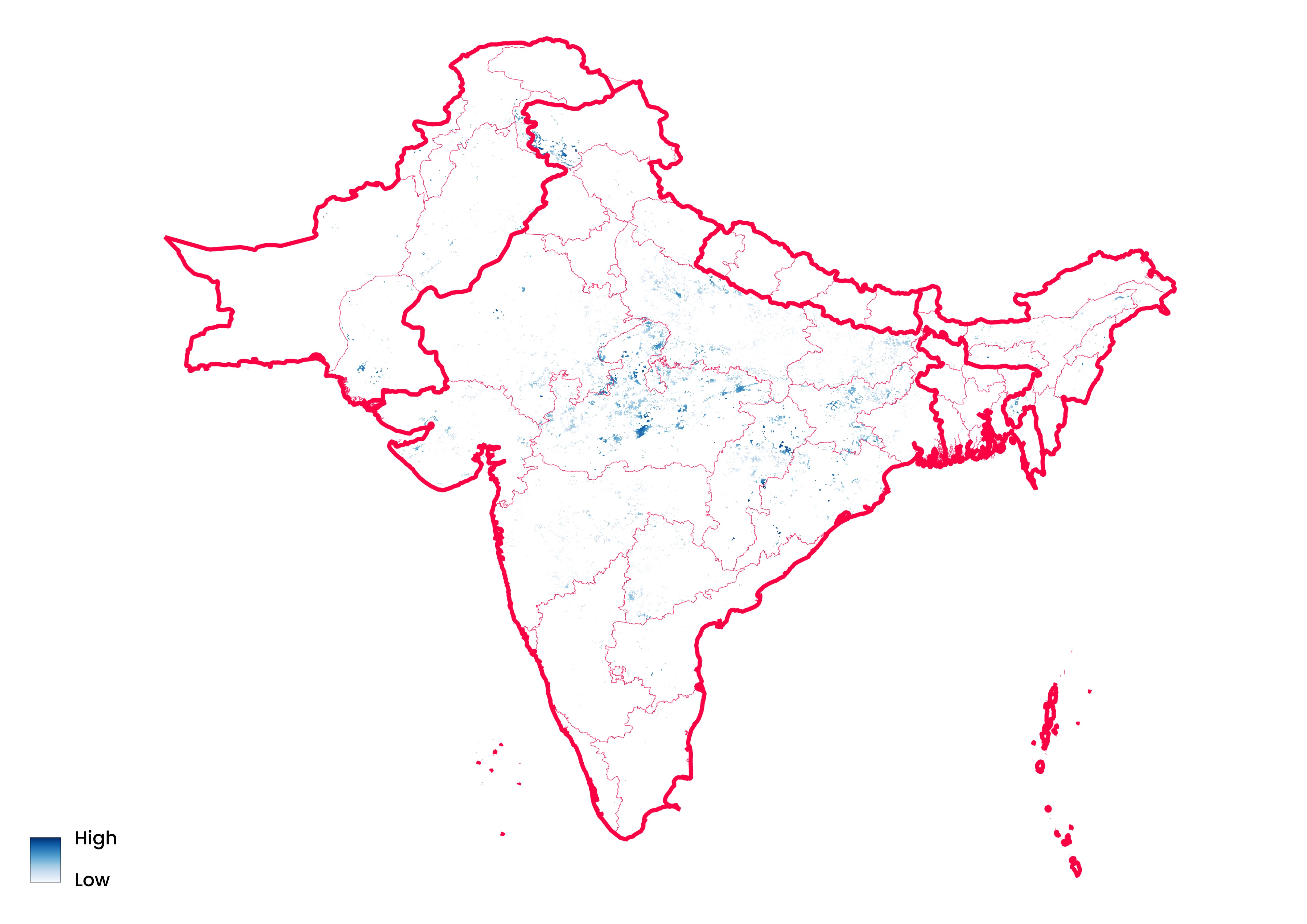} & \includegraphics[width=0.3\textwidth, trim={0.8cm 0 2cm 0},clip]{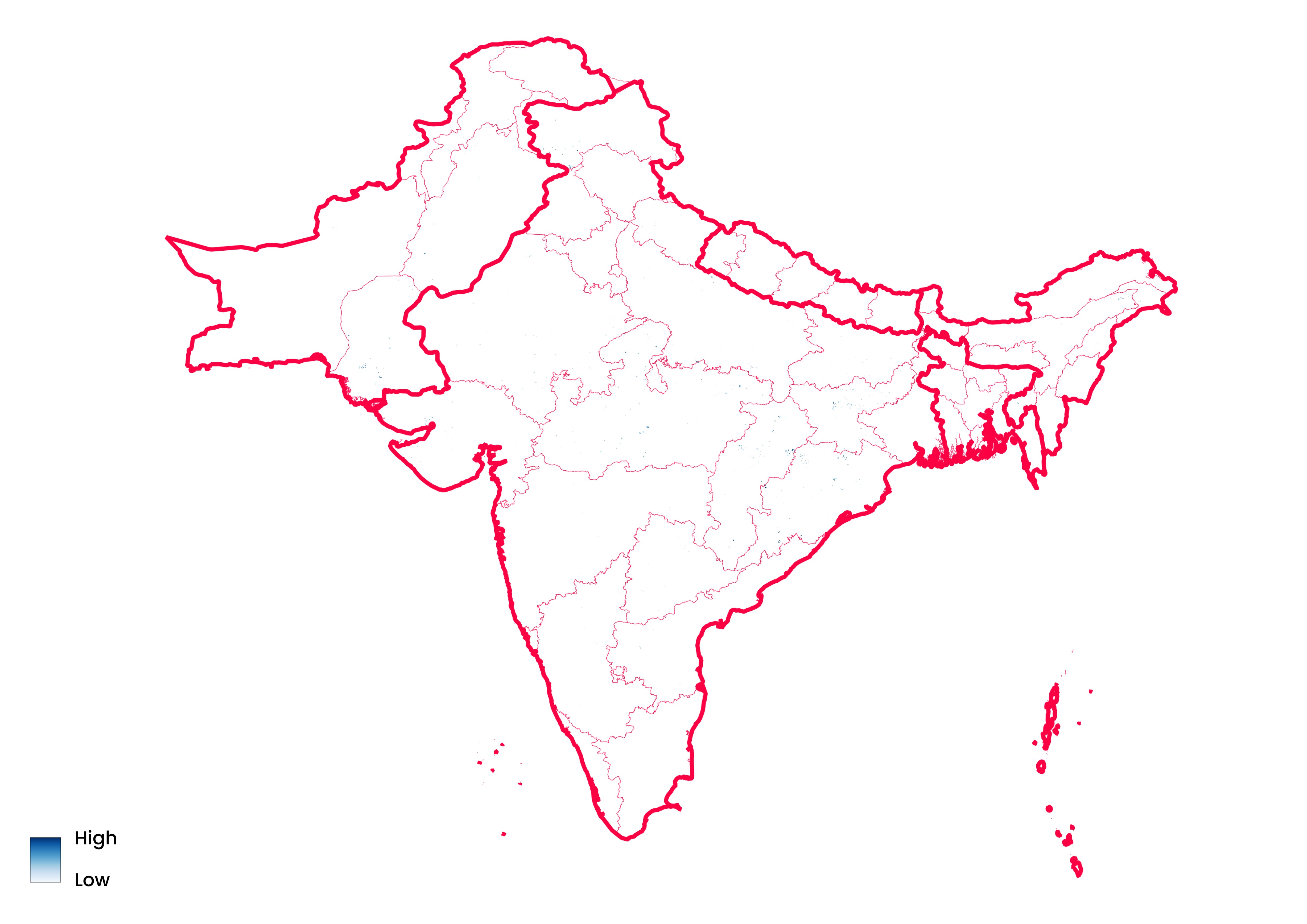} \\
    (a) & (b) & (c) \\
    \end{tabular}
    \caption{Under-served remote areas with (a): Drive time to nearest hospital/clinic is greater than 30 minutes and nighttime light intensity is between 0 and 10; (b): Drive time to nearest hospital/clinic is greater than 30 minutes and nighttime light intensity is between 10 and 20; (c): Drive time to nearest hospital/clinic is greater than 30 minutes and nighttime light intensity is between 20 and 30;}
    \label{fig:findings}
\end{figure}

\section{Results and Evaluation}
\subsection{Analysis}
The analysis is done in multiple steps in order to determine which regions have the greatest need for healthcare facilities and to quantify that need: :
\begin{enumerate}
    \item Quantifying Need Per Pixel: Based on accessibility and population density, a need score is assigned to each region (or pixel). In highly populated locations with limited access to healthcare, the need score is greater (See Fig.~\ref{fig:world} (a) \& (b)).
    \item Aggregating Need Over a Region: A regional need score is produced by adding up the requirements within a 25-kilometer radius. Rather than focusing only on individual pixels, this aggregation aids in identifying larger areas of need (See Fig.~\ref{fig:world} (c)).
    \item Identifying High-Need Regions: By calculating the 99th percentile of the regional need ratings, the areas with the largest requirements are found. The top 1\% of regions with the greatest demand for healthcare facilities are highlighted by applying this criterion. 
    \item Selecting Target Pixels: Target pixels are those that have a need score higher than the ${99}^{th}$ percentile. These are the places with the greatest needs.
    \item Identifying South Asian Regions: Pixels with a need score greater than the 99th percentile are considered targets. These are the areas that require the most assistance.

\end{enumerate}

The final step involves visualizing the results to communicate the findings effectively:

\subsection{Setting Map Center and Displaying Layers} 
A particular region serves as the focal point of the map, and the results are presented through a variety of layers. Among these layers are the following: 
\begin{itemize}
    \item Travel time to healthcare facilities (Refer to Fig.~\ref{fig:datasets}: (a))
   \item Population density (Refer to Fig.~\ref{fig:datasets}: (b))
   \item Nighttime lights intensity for whole world is shown in Fig.~\ref{fig:datasets}: (c)) and we are focusing on South Asia in this study (see Fig.~\ref{fig:ntl}).
   \item Areas with poor healthcare access (Refer to Fig.~\ref{fig:world}: (a))
   \item Densely populated areas with poor healthcare access (Refer to Fig.~\ref{fig:world}: (b))
   \item Regional need scores (Refer to Fig.~\ref{fig:world}: (c))
   \item South Asian regions with the highest need (Refer to Fig.~\ref{fig:SouthAsiaunderserved}: (a, b, c))
\end{itemize}

The regions of the world and South Asia that will benefit most from the construction of new health facilities are clearly identified and illustrated in  Figure \ref{fig:world} and Figure \ref{fig:SouthAsiaunderserved}, providing important information for stakeholders and policy makers. 


\section{Summary of Findings}
This study used three open-source datasets (Accessibility to Healthcare 2019, the Gridded Population of the World Version 4 (GPWv4), Revision 11, and VIIRS Nighttime Day/Night Annual Band Composites V2.2—to map) to assess health disparities on a worldwide scale specifically in South Asia  (see Fig.~\ref{fig:findings}). Our research identifies areas where healthcare facilities are most urgently required by combining data on travel time to medical facilities, population density, and nighttime light intensity.
The findings underline the need of addressing health inequities, particularly in densely populated areas
with limited access to healthcare. We identify the 1\% of regions with the most health services by combining the needs scores calculation with the larger regions. This method ensures that treatments are directed where they are most needed, leading to a more efficient allocation of resources. The production of accurate, up-to-date and publicly available information has been facilitated by advanced technologies such as data science and Earth observation. These technologies not only help us better understand health disparities, but also provide a scalable, long-term solution for continuous development and monitoring. Our data-driven strategy can help reduce health disparities by enabling targeted treatment, especially in resource-constrained areas.


\section{Strategies for Reducing Health Disparities}
Health disparities between people and regions are the result of a complex interaction of social, economic and environmental variables. Identifying these gaps requires a multi-modal approach that includes data-driven insights and community-based solutions. We will review the following strategies based on our proposed strategy to identify where health disparities exist using advanced data analytics and ground observation technology: 

\subsection{Targeted Resource Allocation}
Make use of the high-need areas that have been selected as focus centers for resource distribution. These areas have been identified by data analysis of socioeconomic factors, population density, and healthcare facility accessibility. Governments and healthcare groups can successfully lessen inequities in access to healthcare by allocating resources, such as medical staff, infrastructure, and health education initiatives, to these regions.

\subsection{Addressing Socioeconomic Disparities with NTL Data}
Statistics on nighttime lights (NTLs) can be used to determine how differently affluent people receive healthcare. Examining NTL data reveals disparities in the availability and standard of healthcare facilities according to socioeconomic level, especially in areas such as South Asia where health systems may prioritize the well-off over the underprivileged. Authorities can determine whether locations are devoid of access to high-quality healthcare for poor populations by superimposing NTL data with socioeconomic and health facility maps. This understanding can direct focused initiatives and legislative modifications to enhance impoverished populations' access to and quality of healthcare\cite{toharudin2024investigating}.

\subsection{Leveraging Technology and Telemedicine}
By using telemedicine and technology, we may provide healthcare services to underserved and rural places. By providing consultations, diagnosis, and treatment choices without requiring in-person travel, telemedicine helps close the gap that exists between patients and healthcare practitioners. This strategy can greatly improve access to healthcare and is especially helpful in areas with a weak healthcare system.

\subsection{Encouraging Cross-Sector Collaboration}
To address the socioeconomic determinants of health, we can encourage cooperation across the housing, transportation, education, and healthcare sectors. Since larger social and economic injustices frequently serve as the foundation for health disparities, a concerted effort including several sectors can have a more comprehensive and long-lasting effect on community health.

\subsection{Regulation and Law Enforcement Actions}
Encourage the adoption of evidence-based policies that advance fair access to and use of healthcare services. Regulations that address health inequities should be developed and put into effect by policymakers using data-driven insights. This entails making investments in healthcare infrastructure in places with a high need, helping low-income people pay for healthcare, and making sure that everyone has access to basic medical services.

\section{Policy Framework, Shared Challenges and Joint Actions proposal}
\subsection{Policy Framework}
Primary healthcare is essential to efficient healthcare systems, and developing effective augmentation methods requires a thorough grasp of national health systems. A thorough understanding of the current health infrastructure is necessary to start making significant improvements, and integrating primary healthcare into health system reorganization is essential for effective reforms. One of the most important goals is Universal Health Coverage (UHC), which guarantees fair access to necessary medical treatment without financial hardship. Despite its importance, politicians frequently fail to recognize the specific contributions that primary healthcare makes to population health, which results in inadequate investment in facilities and placing a low priority on community-based specialty training.
Recognized as a cost-effective and low-risk strategy, strengthening primary healthcare aims to achieve universal population coverage by 2030, emphasizing the need for regional collaboration and advocacy. Such initiatives promote prudent investment in community health services and professional training within primary healthcare settings, thereby enhancing healthcare accessibility and supporting UHC goals.
The need to strengthen primary health care is underlined by global cooperation, which is particularly important in regions such as South Asia, which includes Bangladesh, India, Nepal, Pakistan and Sri Lanka - a quarter of the world's population. Given the shared natural resources and increased vulnerability to the effects of climate change, harmonization of health policies and promotion of cooperation between South Asian countries are crucial. Through collaboration, evidence-based solutions can be found, effective health policies can be designed, and initiatives tailored to regional needs can be developed. Collaboration enables innovation in health financing, fighting non-communicable diseases and overcoming shared challenges 
South Asian countries participate in the World Health Organization's (WHO) offices for the South-East Asia Region (SEARO) and the Eastern Mediterranean Region (EMRO), advocating for regional Universal Health Coverage (UHC) strategies. Through the use of resources and knowledge from around the globe, WHO initiatives—which emphasize the significance of primary healthcare in achieving long-term health outcomes—support regional collaboration. This cooperative endeavor aims to enhance climate resilience, promote the welfare of South Asia's sizable populace, and enhance health results. 
\subsection{Shared challenges}
The following list of common issues that South Asian nations face calls for immediate attention and coordinated response from the region's decision-makers:
\begin{itemize}
    \item Population Growth: Over the past few decades, South Asia's population has grown exponentially. India, Pakistan, and Bangladesh are among the world's most populous nations. According to UN estimates, the population of the region will reach astonishing numbers by 2050 if the current growth rate is allowed to continue unchecked. Numerous factors, such as high birth rates, limited access to family planning services, cultural norms, and religious beliefs, are blamed for this growth. . It is necessary to take proactive measures to lower the overall fertility rate and raise the prevalence of contraception in order to combat population growth. Primary healthcare systems must be strengthened in light of demographic changes. This development is hampered, nevertheless, by low funding, inadequate planning, and a lack of knowledge about the role that primary healthcare plays in population health.  
    \item Inequitable Access and Urban-Rural Disparities: In South Asian countries, there are notable geographic differences in the coverage of provinces, districts, and urban-rural areas. Research indicates that populations with lower incomes are probably less healthy, less nourished, less immunized, and less likely to use family planning. Achieving fair access to healthcare requires addressing these disparities. 
    \item Lack of Private Sector Regulation: In South Asia, the private sector is frequently not sufficiently regulated, which causes disparities and inefficiencies in the provision of healthcare. Improving healthcare access and quality requires enforcing compliance and fortifying regulatory frameworks.  
    \item Healthcare Workforce:The shortage of healthcare workers is a major issue for many South Asian nations, especially in rural and underdeveloped areas. Despite efforts to increase the proportion of physicians, dentists, nurses, and paramedics to the general population, disparities and shortages continue to affect the region's overall healthcare system.

    \item Climate Crisis: One of the areas most impacted by climate change is South Asia. Droughts, heat waves, unpredictable rainfall, glacier melting, heat waves, and other extreme weather events have caused massive economic losses and presented significant difficulties for both the populace and governments. Morbidity and mortality are higher in children, the elderly, women, and the homeless, especially those who are impoverished and suffering from diseases. To address the climate crisis and lessen its effects on vulnerable populations, international cooperation and comprehensive policies are needed.

    \item Weak Healthcare Ecosystem: Developing a strong healthcare ecosystem is a major challenge for South Asia, which includes nations like India, Pakistan, Bangladesh, Nepal, Sri Lanka, Bhutan, and the Maldives. Even though the region's health indicators have improved, there are still several systemic problems that prevent healthcare services from being delivered effectively in the area. Inadequate funding, undeveloped supply chains for medical equipment, a lack of pharmaceutical company involvement in the management of non-communicable diseases (NCDs), and a dearth of healthcare facility developers are major factors in the weak healthcare ecosystem.  
    \begin{itemize}
        \item Healthcare Financing:Throughout South Asia, there is a severe lack of investment in the health sector. The amount of money allotted by the government to healthcare is frequently insufficient, usually much less than the 5 percent of GDP that is advised to be used in order to significantly advance Universal Health Coverage (UHC). For instance, the amount spent on healthcare in nations like Bangladesh and Pakistan is only a small portion of GDP—roughly 2-3 percent. Due to lack of funding, there are not as many resources available for developing infrastructure, training the healthcare workforce, or buying necessary medical supplies. Many health initiatives lack sufficient funding due to the insufficiency of donor funding and the private sector.  
        \item Medical Equipment Suppliers and Manufacturers: South Asia's medical equipment supply chain is undeveloped and dispersed. Due to a lack of manufacturing capacity locally, a significant amount of medical equipment and devices are imported. This reliance raises expenses and causes delays in the supply of essential equipment. In addition, variations in regulatory requirements for medical devices lead to discrepancies in terms of both safety and quality. Improving the availability of medical equipment requires stricter regulatory frameworks and stronger local manufacturing capabilities.  
        \item Pharmaceutical Companies and NCD Medicines:South Asia is experiencing a surge in non-communicable diseases (NCDs), including diabetes, hypertension, and cardiovascular disorders. But the pharmaceutical industry hasn't been able to meet this increasing demand. Local pharmaceutical firms frequently place a greater emphasis on generic drugs than on cutting-edge NCD therapies. There is a shortage of reasonably priced NCD drugs, especially in rural and low-income areas. Improving regulatory supervision and pharmaceutical companies' ability to create and market NCD medications are essential for better NCD management. 
        \item Healthcare Facilities Developers: Hospitals, clinics, and diagnostic centers are among the healthcare facilities that have not developed equally throughout South Asia. While access to these services may be somewhat better in urban areas, rural and isolated areas are still woefully underprivileged. Funding shortages, land acquisition concerns, and bureaucratic roadblocks frequently impede investments in healthcare infrastructure. Furthermore, the upkeep and modernization of current facilities are commonly disregarded, which results in declining service quality. Promoting public-private partnerships and optimizing regulatory procedures can expedite the establishment of healthcare facilities and enhance the availability of superior healthcare services. 
    \end{itemize}

\subsection{Joint Actions for Strengthening Healthcare Ecosystems in South Asia}
The healthcare ecosystems in South Asian nations need to be strengthened, and this will require a concerted and cooperative effort. Together, a data-driven strategy, policymakers' alignment, resource sharing, regional coordination, quick response capabilities, and a heavy emphasis on data science will enhance the region's resilience, healthcare delivery, and results. Through cooperation and shared commitment, South Asia can achieve a more robust and equitable healthcare system for all of its citizens. By implementing these coordinated actions, South Asian countries can significantly strengthen their healthcare ecosystems. 
\begin{itemize}
    \item Data-Driven Approach
    \begin{itemize}
        \item Establish Regional Health Data Repositories: Construct a single, easily accessible repository for health data that can be used by all of South Asia's nations to gather, store, and process health data. Data on the prevalence of diseases, healthcare resources, patient outcomes, and other pertinent metrics ought to be included in this repository. 
        \item Implement Standardized Health Information Systems: To guarantee data consistency and interoperability, adopt standardized health information systems throughout the region. This will make it easier for health data from different nations to be shared and compared. 
        \item Utilize Advanced Analytics: Employ advanced data analytics and machine learning techniques to identify trends, predict outbreaks, and optimize resource allocation. This will enable a proactive rather than reactive approach to healthcare management.

    \end{itemize}
    \item Align Policymakers and Local Bodies
\begin{itemize}
    \item Regular Inter-Governmental Health Summits: Arrange for policymakers from South Asian nations to convene on a regular basis to tackle shared healthcare issues, exchange optimal methodologies, and harmonize on regional health policies.
    \item Collaborative Policy Frameworks: Develop and implement collaborative policy frameworks that address cross-border health issues such as infectious disease control, healthcare financing, and workforce development. 
    \item Engage Local Authorities: Make sure local organizations actively participate in the creation and execution of policies, taking into account their particular perspectives and practical experience.  
\end{itemize}
    \item Sharing Resources and Transferability
    \begin{itemize}
        \item Resource Sharing Agreements: Create official contracts for the sharing of essential healthcare resources like drugs, medical supplies, and knowledge. This will facilitate access to necessary services and assist in addressing shortages. 
        \item Cross-Border Healthcare Training Programs: Implement cross-border training programs for healthcare professionals to build capacity and ensure the transferability of skills and knowledge across the region.

        \item Telemedicine and Remote Consultation Services: Increase the availability of telemedicine services to facilitate the sharing of resources and offer online consultations, particularly in underprivileged areas. 
    \end{itemize}
    \item Coordination Among Regions
    \begin{itemize}
        \item 	Regional Health Coordination Bodies: Set up regional health coordination bodies to facilitate cooperation and coordination among South Asian countries. These bodies can oversee joint initiatives, monitor progress, and address emerging health challenges. 
        \item Joint Research and Development Initiatives: Encourage collaborative R\&D projects centered around non-communicable diseases, infectious diseases, and technological advancements in healthcare. 
        \item Emergency Response Coordination: To effectively manage health crises such as pandemics, natural disasters, and other emergencies, develop a coordinated emergency response framework. 
    \end{itemize}
    \item Rapid Action Teams
    \begin{itemize}
        \item Establish Regional Rapid Response Teams: Form multidisciplinary rapid response teams that can be deployed quickly to any South Asian country facing a health crisis. These teams should include medical professionals, epidemiologists, logistics experts, and communication specialists.

        \item Regular Drills and Simulations:Hold frequent drills and simulations to make sure rapid response teams are capable of cross-border operations and are ready for a variety of scenarios.  
        \item Real-Time Communication Channels: Set up real-time communication channels for rapid action teams to coordinate efforts, share information, and mobilize resources swiftly.

    \end{itemize}
    \item Data Scientists Team
    \begin{itemize}
        \item Regional Data Science Consortium: Create a consortium of data scientists from South Asian countries to collaborate on health data analysis, predictive modeling, and decision support systems.
        \item Training and Capacity Building: Fund educational initiatives to create a strong pool of data scientists with backgrounds in epidemiology and health informatics. 
        \item Data-Driven Decision Support Tools: Create and implement tools for decision-making that use data analytics to guide the allocation of resources, clinical procedures, and policy decisions 
    \end{itemize}
    
\end{itemize}

\end{itemize}

          

\bibliographystyle{splncs04}
\bibliography{refs}

\newpage
\begin{appendix}
\section{AccessMod}\label{accessmod}
AccessMod\footnote{\url{https://www.accessmod.org/}} is a developer's software that has a limitation to Linux operating system only, it requires multiple datasets in different formats, having lot more complexity to handle the projection errors of other data files on the Base DEM. It requires the same projection of all data files having different extensions i.e: .shp, .shx,  .dbf, .tif and .csv. AccessMod is hard to compute `Access to facilities' having such projection problems. It cannot localize health disparities in underserved remote areas.

\begin{table}
    \centering
    \begin{tabular}{cc}
    \includegraphics[scale=0.305]{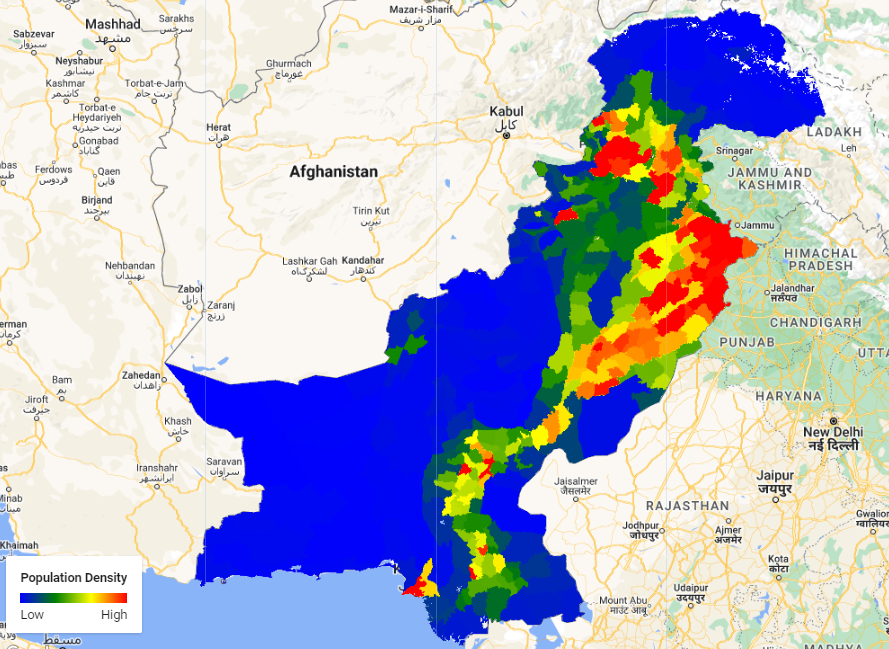}   
         & \includegraphics[scale=0.3]{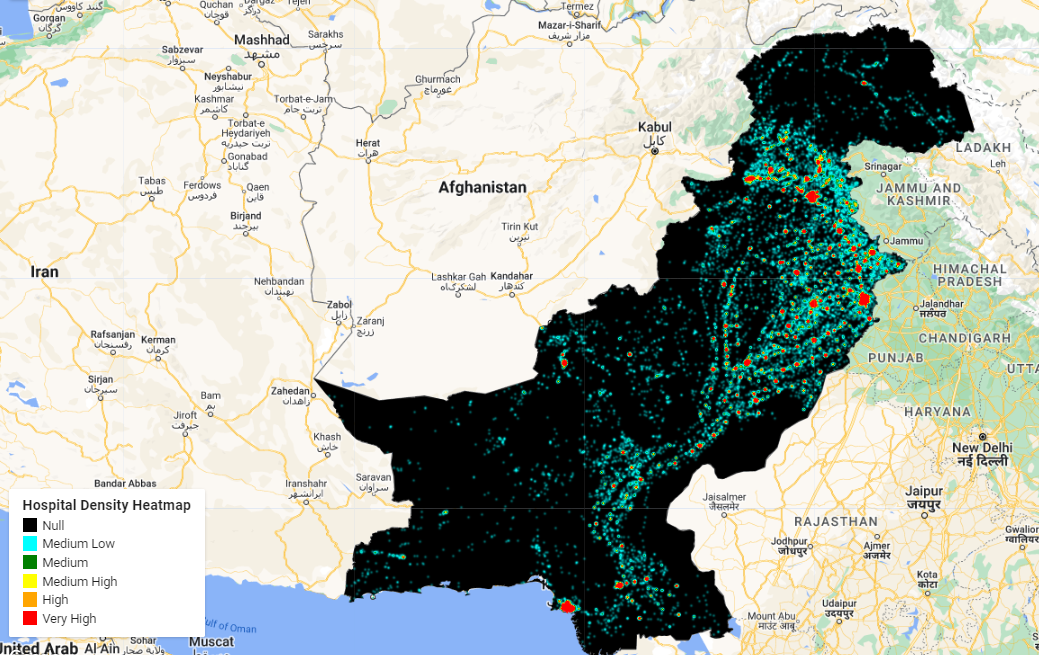}\\
         (a) & (b)
    \end{tabular}
    \caption{(a): Population density and (b): health disparities in Pakistan, using AccessMod.}
    \label{tab:accessmod}
\end{table}

\end{appendix}

\end{document}